\documentclass[preprint,12pt]{elsarticle}

\usepackage{amssymb}
\usepackage{color} 
\usepackage{subcaption}
\usepackage{abbrev}
\usepackage{siunitx}
\usepackage{fancyvrb}
\usepackage[T1]{fontenc}
\usepackage{natbib}
\usepackage{ulem}
\usepackage{url}

\definecolor{mbul}{rgb}{0.5,0.25,1} 

\usepackage{lineno}

\journal{NIM A}

\begin{document}

\begin{frontmatter}


\author[a,b]{Rakhee Kushwah\corref{cor1}}
\author[a,b]{Nirmal K. Iyer}
\author[a,b]{M\'ozsi Kiss}
\author[a,b]{Theodor A. Stana}
\author[a,b]{Mark Pearce}

\cortext[cor1]{Corresponding author\ead{rakhee@kth.se}}
\address[a]{KTH Royal Institute of Technology, Department of Physics,\\106 91 Stockholm, Sweden}
\address[b]{The Oskar Klein Centre for Cosmoparticle Physics, AlbaNova University Centre,\\106 91 Stockholm, Sweden}

\title{A Compton polarimeter using scintillators read out with MPPCs through Citiroc ASIC}


\author{}

\address{}

\begin{abstract}
In recent years, a number of purpose-built scintillator-based polarimeters have studied bright astronomical sources for the first time in the hard X-ray band (tens to hundreds of keV). The addition of polarimetry can help data interpretation by resolving model-dependent degeneracies. The typical instrument approach is that incident X-rays scatter off a plastic scintillator into an adjacent scintillator cell. In all missions to date, the scintillators are read out using traditional vacuum tube photo-multipliers (PMTs). The advent of solid-state PMTs (“silicon PM” or “MPPC”) is attractive for space-based instruments since the devices are compact, robust and require a low bias voltage. We have characterised the plastic scintillator, EJ-248M, optically coupled to a multi-pixel photon counter (MPPC) and read out with the Citiroc ASIC. A light-yield of 1.6 photoelectrons/keV has been obtained, with a low energy detection threshold of $\lesssim$5~keV at room temperature. We have also constructed an MPPC-based polarimeter-demonstrator in order to investigate the feasibility of such an approach for future instruments. Incident X-rays scatter from a plastic-scintillator bar to surrounding cerium-doped GAGG (Gadolinium Aluminium Gallium Garnet) scintillators yielding time-coincident signals in the scintillators. We have determined the polarimetric response of this set-up using both unpolarised and polarised $\sim$50~keV X-rays. We observe a clear asymmetry in the GAGG counting rates for the polarised beam. The low-energy detection threshold in the plastic scintillator can be further reduced using a coincidence technique. The demonstrated polarimeter design shows promise as a space-based Compton polarimeter and we discuss ways in which our polarimeter can be adapted for such a mission. 
\end{abstract}

\begin{keyword}


Gamma/X-ray detector, scintillator, MPPC, GAGG, plastic scintillator, Compton polarimetry
\end{keyword}

\end{frontmatter}


\section{Introduction}
\label{sec:intro}

In astrophysics, X-ray polarimetry offers a systematically different observation methodology compared to the usual imaging, spectroscopy and timing approaches \cite{polsci1,polsci2}. Instruments with polarimetric capability allow source emission to be characterised with two additional observables, the linear polarisation fraction (PF) and the linear polarisation angle (PA). The addition of these quantities stands to disentangle physical and geometric effects, thereby resolving model-dependent degeneracies which often complicate the interpretation of observations. Polarisation is a positive-definite quantity and should be measured using purpose-built instruments which are extensively calibrated with both polarised and unpolarised beams before launch. 

Such instruments have provided reliable polarisation information on bright sources like the the Crab \cite{crab}, Cygnus X-1 \cite{cygnus} and gamma-ray bursts \cite{gap, polarsci} through the study of Compton scattering of hard X-ray ($\sim$30-500 keV) photons in a scintillator array. X-rays scatter preferentially perpendicularly to the polarisation vector. The distribution of azimuthal scattering angles ('modulation curve') has a phase (amplitude) which defines PA (PF). These two quantities can also be derived from an unbinned Stokes parameter analysis \cite{stokes}.

In such scintillator arrays, plastic scintillator is generally used as the scatterer since it has a low atomic number and presents a high probability for Compton scattering. The material is also advantageous as its behaviour in the (near-)space environment is very well documented. In single-phase instruments, e.g. PoGO+ \cite{pogo} and POLAR \cite{polar}, only plastic scintillator material is used for detecting both the scattering and the photoelectric absorption. This allows a low-cost and low-mass polarisation target to be realised. However for higher energies, the photoabsorption cross-section in plastic is not as favourable as in higher atomic number materials such as CsI(Tl), as used e.g. in GAP \cite{gap}. Such two-phase instruments therefore stand to offer a superior modulation response as well as good energy resolution. For the aforementioned missions, the scintillators are read out using photomultiplier tubes (PMTs).

The multi-pixel photon counter (MPPC)\footnote{Also known as silicon photo-multiplier.}, is a solid-state photon-counting device with typically 10$^3$-10$^5$ avalanche photodiode pixels operated in Geiger-mode above the breakdown voltage. The output signal is formed as the analogue sum of the triggered pixels (proportional to number of detected photons). The gain and intrinsic signal timing characteristics are comparable to those of a PMT. Compared to a PMT, the MPPC has many advantages, e.g. lower operation voltage ($\sim$50 V), insensitivity to magnetic fields, low mass ($\sim$1 g for single devices), and a compact and robust design. Although the sensitive area for single MPPC devices (typically $\sim$5$\times$5 mm$^2$) is significantly smaller than for PMTs (typically, at least several cm$^2$), this can be addressed by using a tiled array of single \mbox{MPPCs} depending on the desired instrument geometry. 

For application to a plastic-scintillator-based Compton polarimeter, the largest challenge is posed by the nature of the signal from a photon which Compton scatters in plastic. For an instrument with a representative lower energy limit of 50 keV, a photon scattering through a polar angle of 90$^\circ$ will only deposit $\sim$5 keV. The effects of non-linear light-yield (quenching) in the scintillator will further suppress the conversion of deposited energy into the scintillation light signal. A disadvantage of the MPPC is the high dark count rate. If required, this can be alleviated by operating devices below room temperature, e.g. using Peltier effect coolers.  For polarimetric applications, the effect of high dark count rate is naturally reduced by the short ($\sim$100 ns) coincidence window used when identifying simultaneous scatter/absorption events in the scintillator array \cite{naka2016,gap2019}. Another issue is that the MPPC gain is temperature-dependent which requires the bias voltage to be regulated in a temperature-controlled feedback loop. Many commercial MPPC power supplies provide such functionality. Finally, and significantly, there is little experience in using MPPCs in the space environment, and radiation tolerance is a concern. Interactions with hadrons are known to cause an increase in the dark current and reduce gain \cite{noisesipm,radhardsipm}. These effects will degrade the low energy response of a polarimeter based on MPPC readout. Mitigating the effects of radiation damage in MPPCs is currently an active area of research, and methods like thermal annealing have been shown to be effective in reducing the increase in dark currents \cite{damagesipm,neutronsipm}.

\begin{figure}[h]
\centering 
\includegraphics[width=8cm, height=6cm]{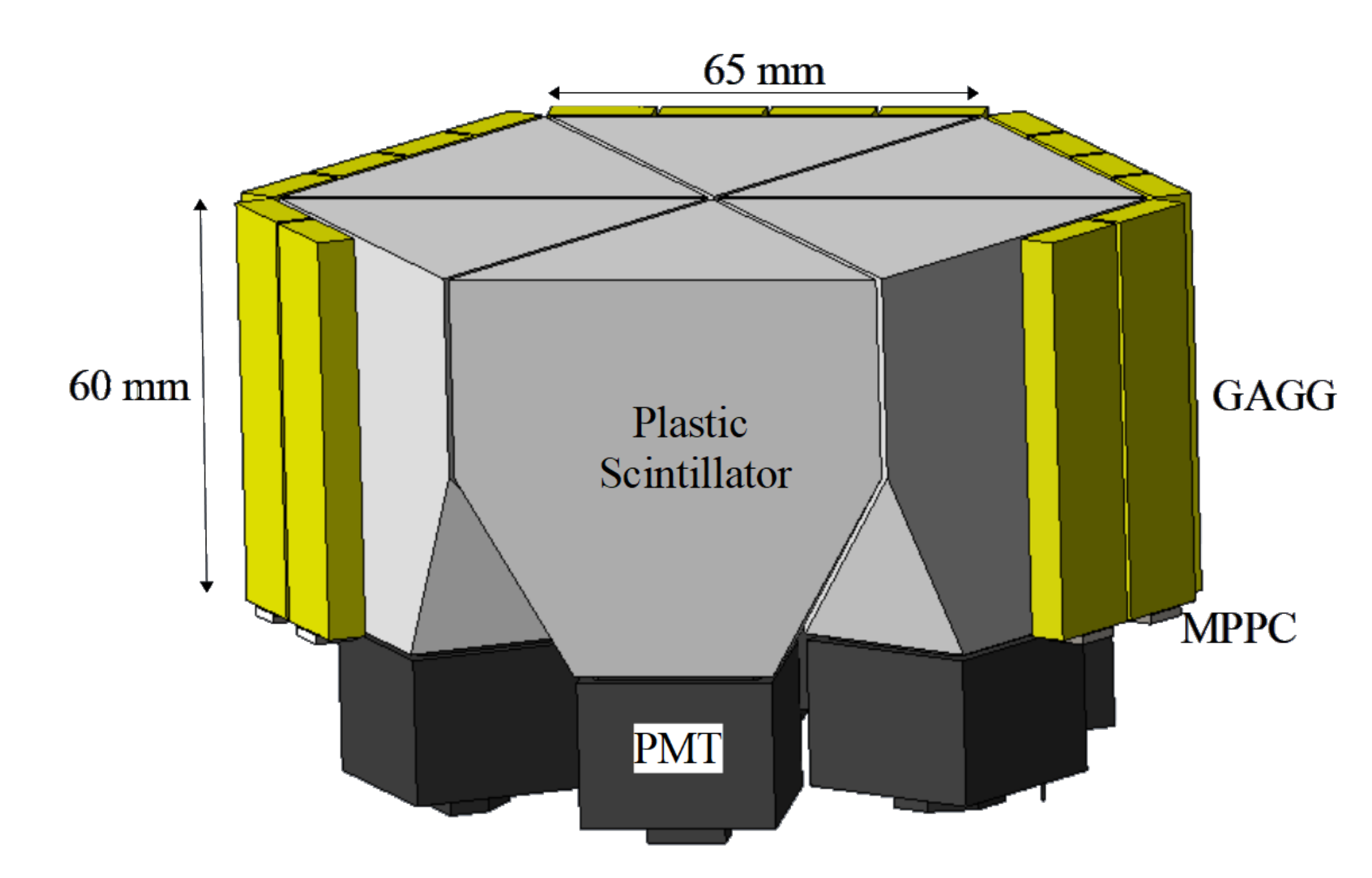}
\caption{\label{fig:geo} The geometry of the plastic and GAGG scintillators proposed for the SPHiNX mission. A part of the GAGG wall has been hidden to show the plastic scintillator geometry more clearly. These GAGG elements are used in the work presented here.}
\end{figure}

We recently proposed a large field-of-view instrument optimised for gamma-ray burst polarimetry, SPHiNX~\cite{sphinx}, which uses a plastic scintillator scatterer (PMT read-out) combined with a GAGG(Ce) scintillator (Gd$_{3}$Al$_{2}$Ga$_{3}$O$_{12}$(Ce)) with MPPC read-out. One of its detector modules is depicted in Figure~\ref{fig:geo}. Inspired by the design of SPHiNX, in this paper we explore the feasibility for a small satellite large field-of-view polarimeter design using compact/low voltage MPPCs connected to low-power multi-channel application-specific integrated circuits (ASICs). An important focus of our work is to determine the lower energy detection limit achievable using a plastic scintillator with MPPC read-out. The fast MPPC signals required for the coincidence measurements combined with the relatively high terminal capacitance of the MPPC, mean that reading out low-amplitude noise-free signals is potentially problematic. In this work, we explore the use of the Citiroc ASIC for this purpose. Other types of read-out ASICs are being studied elsewhere \cite{siphraASIC,multichASIC}.

The next section details tests of the polarimeter components, i.e. scintillators, MPPCs and the ASIC used to process the MPPC signals. In Section~\ref {sec:xraypolarimetry}, the performance of a prototype polarimeter is reported for both an unpolarised and a polarised $\sim$50 keV beam of photons. We are able to make a clear detection of the modulated counting rate for a polarised beam. These results therefore represent an important step forward compared to previous related work \cite{gap2019,bloser,compass}.

\section{Components of a Compton polarimeter-demonstrator}
\label{sec:Prerequisites}
\begin{figure*}[h]
\centering 
\includegraphics[width=\textwidth]{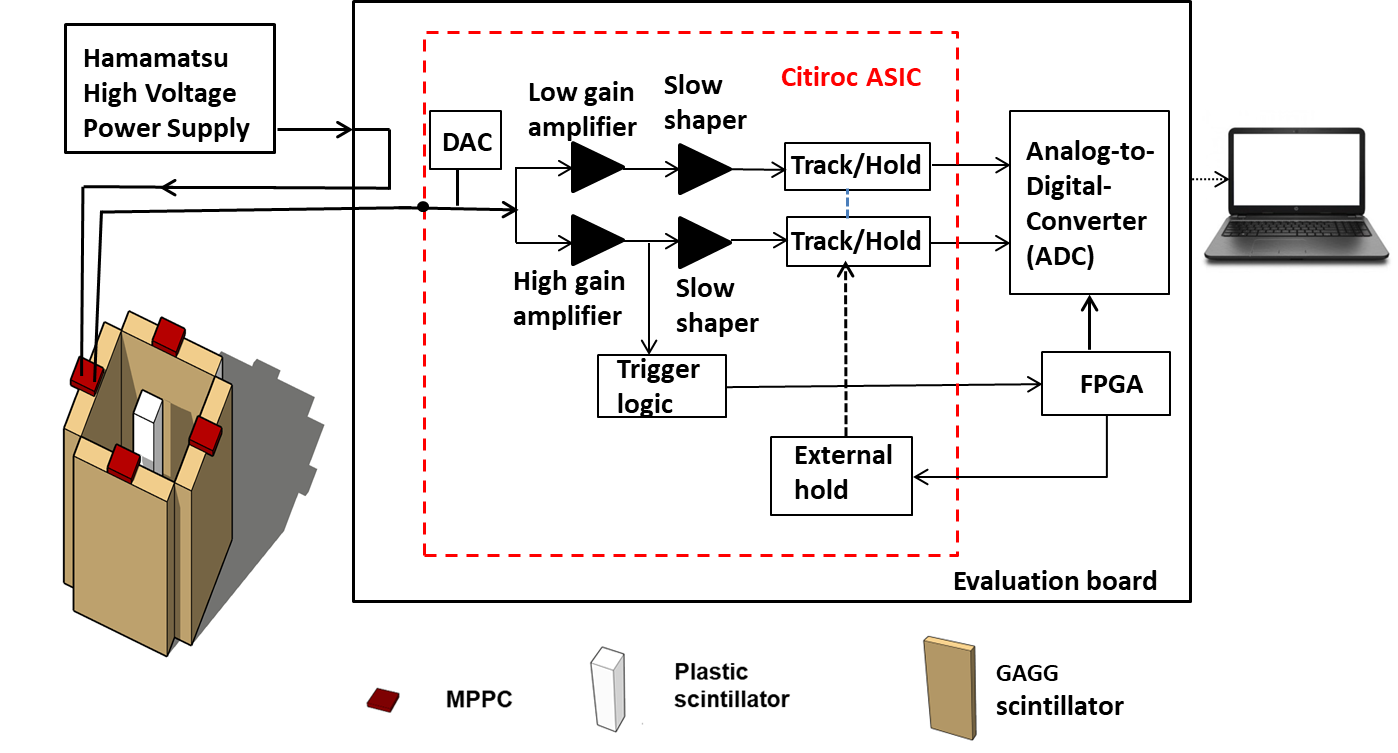}
\caption{\label{fig:schematic} Schematic of a Compton polarimeter-demonstrator. The electronics chain of one of the channels is shown. Each channel has the same architecture. The MPPC on the plastic scintillator is connected underneath and is not visible in the figure.}
\end{figure*}

A schematic representation of the components which are required to build a Compton polarimeter-demonstrator is shown in Figure \ref{fig:schematic}. A bar of plastic scintillator is surrounded by four GAGG scintillators. Each of the scintillators is optically coupled to a respective MPPC. The output signal from each MPPC is fed to individual channels of an ASIC called Citiroc \cite{citichar} manufactured by Weeroc. Presently, the Citiroc is used on a Weeroc evaluation board. The evaluation board also carries an Altera III field-programmable gated array (FPGA) and a 12-bit analog-to-digital converter (ADC) for processing the signals from Citiroc. Processed data is written to a file which is transferred to a PC via a USB cable.

\subsection {Characterisation of the Citiroc}
\label{sec:Citiroc}
The Citiroc 1A is a 32-channel analogue front-end ASIC for MPPC read-out and it accepts positive polarity input signals. Each channel features an input digital-to-analog converter (DAC) for adjusting the bias voltage of an MPPC. The MPPC signal is fed to low gain (LG) and high gain (HG) amplifiers with a gain ratio of 10. A range of gain values is available with each amplifier. Each amplifier is followed by a slow CR(RC)$^2$ shaper (one differentiator stage and two integrator stages) with a shaping time configurable within the range 12.5 ns $-$ 87.5 ns. For our application, the shaping time is configured to 37.5 ns. The internal triggering logic can be driven by either of the LG or HG amplifiers. In the current setup, the HG chain is used for the triggering.  The trigger logic consists of two independent discriminators. One of them (Charge trigger) provides a single trigger output for all 32 channels while the other one (Time trigger) provides a trigger output for each channel. In the present study, the OR-ed time trigger signal is used to feed the external hold signal via the FPGA. The shaped outputs are held for a programmable time duration ($\sim$200 ns) upon receipt of an external signal, thereby allowing coincidence studies. As the Citiroc ASIC was developed for use with ground-based Cherenkov Telescopes, its radiation tolerance must be determined for space missions. We note that a related ASIC, EASIROC, on which the Citiroc design is based \cite{citichar}, has been found to be suitable for use in radiation environments\footnote{\url{https://agenda.infn.it/event/4107/contributions/50274/attachments/35384/}}.

In our study, the Citiroc trigger threshold values are chosen to lie above the pedestal of the ASIC and the dark-noise of the MPPC. The pedestal of the ASIC (at $\sim$960 ADC channels) is due to baseline fluctuations and defines the zero reference of the ADC, dictated by the evaluation board. The MPPC dark-noise consists of multiple photoelectron (p.e.) peaks arising due to thermally generated electrons triggering the corresponding number of MPPC pixels. At ambient operating conditions, no peaks are seen above the 3rd p.e. on any of the tested MPPCs. Thus, the trigger threshold on Citiroc is set above the 3rd p.e. peak for each MPPC. Any signal above a trigger threshold is referred to as a "hit". The output data-file consists of information on each hit for every channel. It is formatted as an ASCII space-separated text file with each row corresponding to a valid hit in the Citiroc. There are three columns per channel viz. LG, HG and hit flag. The LG and HG columns have an ADC value for the corresponding hit. The hit flag is asserted if the signal is above the trigger threshold on that particular channel.

\begin{figure*}[h]
\begin{subfigure}[t]{0.5\linewidth}
\includegraphics[width=\linewidth]{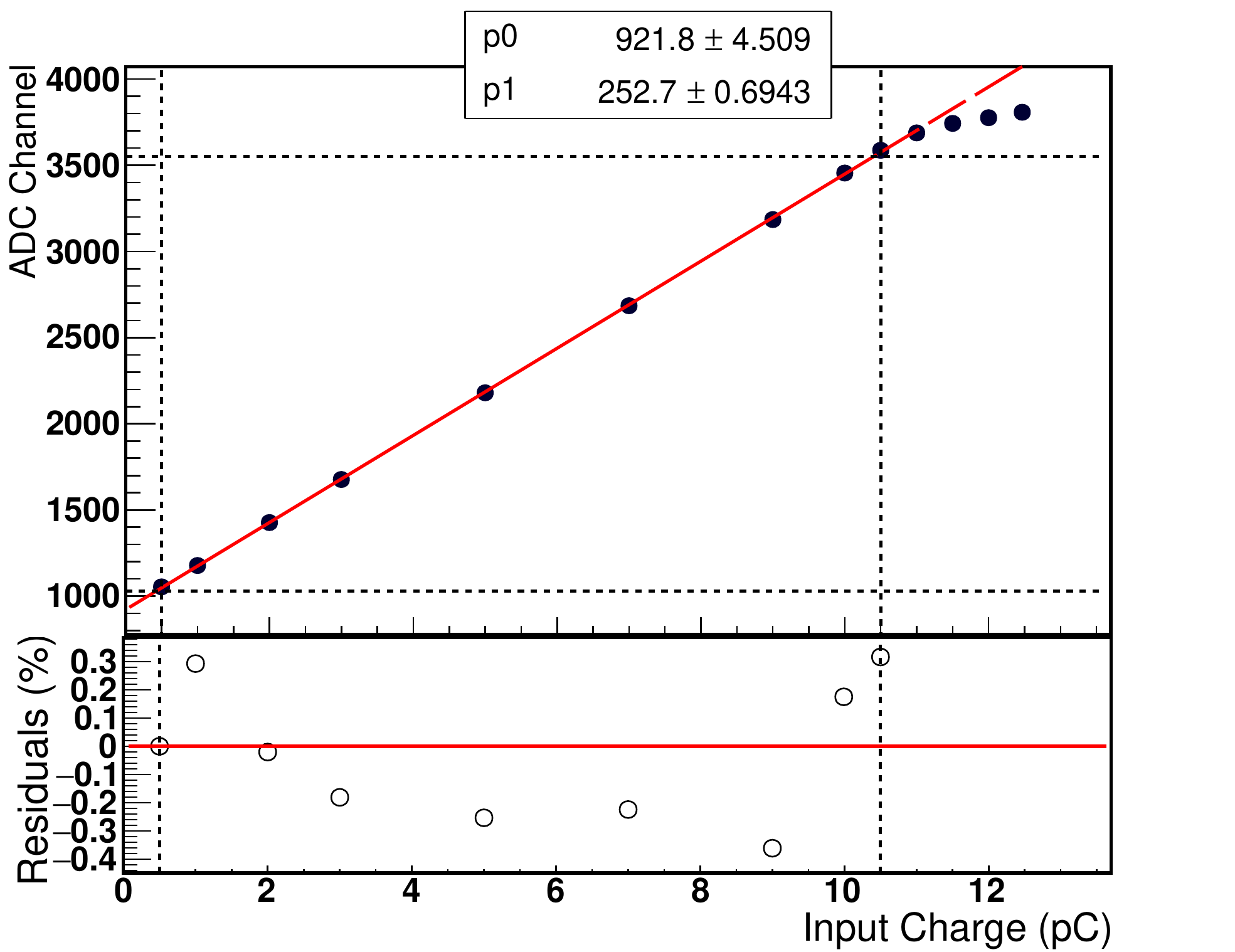}
\caption{\label{fig:LinearityHG}HG amplifier of Channel 0 at gain = 40. \\The charge on the horizontal axis corresponds to \\the energy from a few keV to a few tens of keV.}
\end{subfigure}
\begin{subfigure}[t]{0.5\linewidth}
\includegraphics[width=\linewidth]{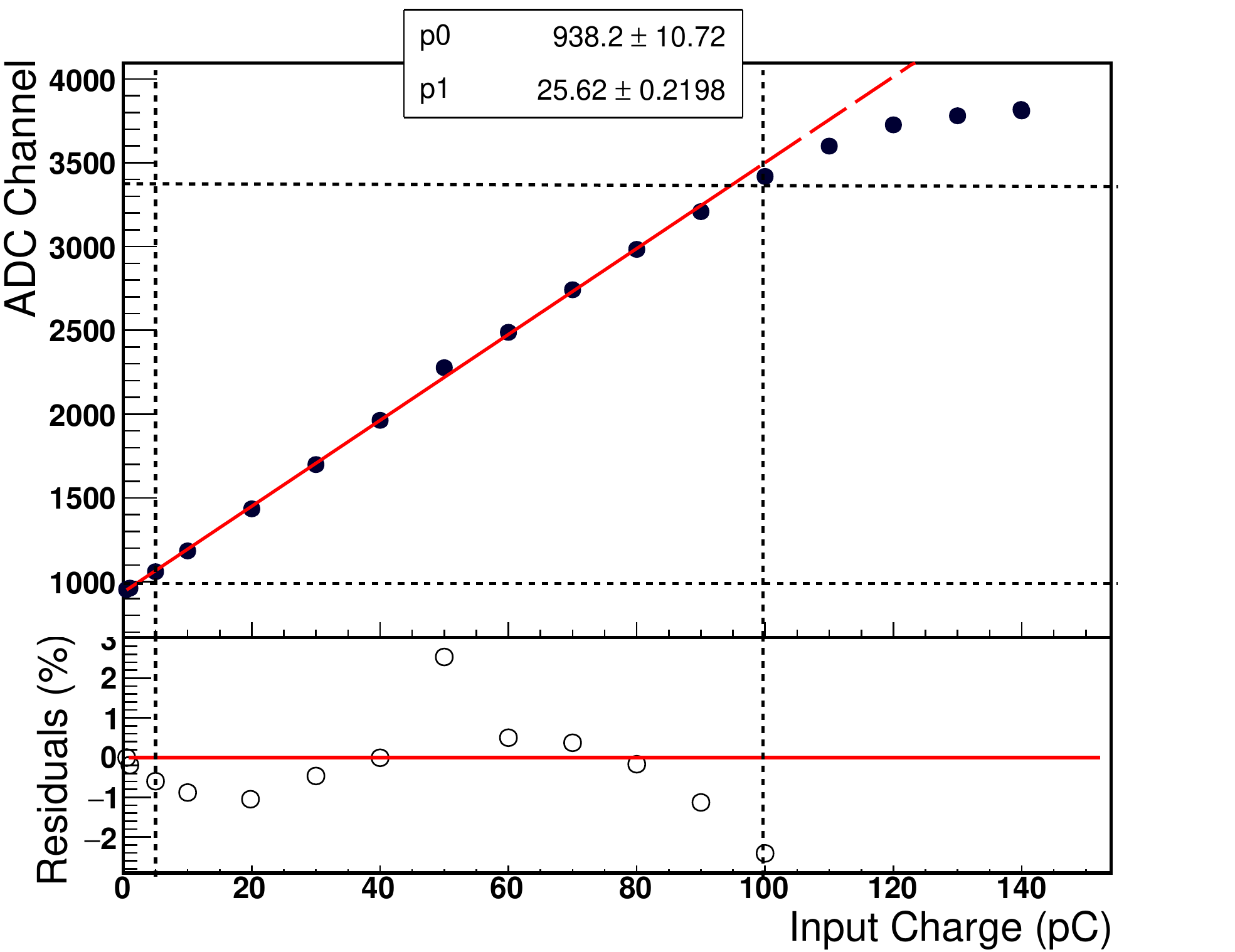}
\caption{\label{fig:LinearityLG}LG amplifier of Channel 0 at gain = 7.5. \\The charge on the horizontal axis corresponds to \\the energy from a few tens of keV to a few hundreds of keV.}
\end{subfigure}%
\caption{Linearity of Citiroc.}
\end{figure*}

Since the Citiroc is used to perform spectroscopy, the linearity of its response is evaluated using a pulse generator, RIGOL DG1062Z. The output pulse from the pulse generator is sent to the Citiroc channel under study via a 100 pF capacitor to convert the voltage pulse into a current pulse. Linearity plots of the HG and LG amplifiers for a single Citiroc channel are shown in Figure~\ref{fig:LinearityHG} and Figure~\ref{fig:LinearityLG} respectively. Measured values are fitted with a linear function, $\mathrm{p0+p1\textit{x}}$, where p0 is the baseline of the ASIC and p1 is the slope. The percentage residuals are also shown in the same figure. Some non-linearity is observed at the higher ADC channels. The vertical dashed lines in the plots represent the linear range of the amplifier. The corresponding ADC channels, shown with horizontal dashed lines, are used for deriving the results presented in this paper. The linearity of HG and LG amplifiers on all 32 channels of Citiroc has been measured at the same gain setting and similar results are obtained.

\subsection{Characterisation of scintillators}
\label{sec:scintillators}
\begin{figure}[h]
\begin{subfigure}[t]{0.45\linewidth}
\includegraphics[width=\linewidth]{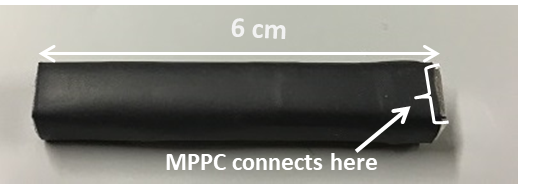}
\caption{\label{fig:Plastic}Plastic scintillator.}
\end{subfigure}
\begin{subfigure}[t]{0.45\linewidth}
\includegraphics[width=\linewidth]{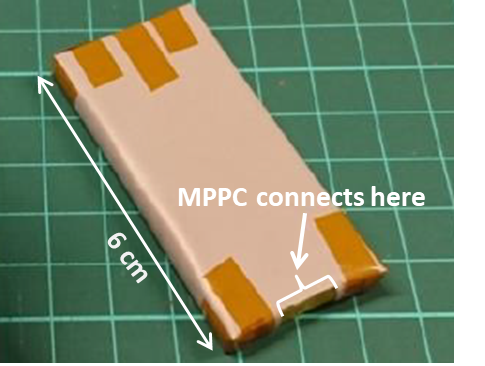}
\caption{\label{fig:GAGG}GAGG scintillator.}
\end{subfigure}%
\caption{\label{fig:plasgagg}Two types of scintillators used to build a polarimeter-demonstrator.}
\end{figure}

The plastic scintillator used for this study is an organic scintillator of type EJ-248M from Eljen Technology. It has dimensions 60 $\times$ 6 $\times$ 6 mm$^{3}$. The scintillator is wrapped with two layers of Vikuiti$^{TM}$ Enhanced Specular Reflector (ESR) material and a single layer of Tedlar$^{\circledR}$, a high-opacity material for light-tightening. It is also covered by a black heat-shrink tube for protection. The wrapped plastic scintillator is shown in \mbox{Figure~\ref {fig:Plastic}}.

GAGG is an inorganic scintillator from the company, C\&A. Compared to other inorganic scintillators, it has advantages like higher density, leading to high stopping power, and high light-yield. It is non-hygroscopic and its light-yield shows a 20\% variation in the temperature range, between \mbox{-20 $-$ 20 $^{\circ}$C} \cite{LYtemp}. As it is a recent development, its use in the space radiation environment is not fully explored \cite{sphinx,gaggactive}. The GAGG used for this work is \mbox{60 $\times$ 27.5 $\times$ 5 mm$^{3}$} in size. It is also wrapped with two layers of ESR and a layer of Tedlar as shown in Figure~\ref{fig:GAGG}. The properties of the plastic and GAGG scintillators are summarised in Table~\ref{table:scintillators}.

A monolithic MPPC, Hamamatsu S13360-6050CS, is used for this study. It has an effective photosensitive area of \mbox{6 $\times$ 6 mm$^{2}$}. The MPPCs are powered using a Hamamatsu high voltage power supply (C11204-01) via the Citiroc evaluation board. The DAC in each corresponding channel of the Citiroc is tuned to make all \mbox{MPPCs} have comparable gains. Biased at the maximum allowed 55 V, the MPPCs are operating at a gain of approximately 6 $\times$ 10$^{5}$. All scintillators are optically coupled to the same type of MPPC using silicone gel, EJ-550. 

The set-up shown in Figure~\ref {fig:schematic} is kept inside a dark-box to prevent signal degradation by ambient light. The same set-up is also used to study the spectral response of the scintillators. Spectra are acquired for each of the scintillators by irradiating with different radioactive sources. All measurements are conducted at room temperature.

\begin{table}[h]
\centering
\caption{Scintillator properties.}
\label{table:scintillators}
\begin{tabular}{|l|c|c|}
\hline
					& {\bf EJ-248M} 	& {\bf GAGG}   \\
\hline
Light output ($\gamma$/MeVee) & 9.2 $\times$ 10$^{3}$  &    5.6 $\times$ 10$^{4}$              \\ 	
Decay time (ns)		 	& 	$\sim$2	&     88             \\  
Density (g/cm$^{3}$) 	&	1.02				& 6.63                  \\ 
Peak wavelength (nm) 	&	425				&       520        \\ 
Refractive index & 1.6 & 1.9 \\	
\hline
\end{tabular}
\end{table}

\subsubsection{Plastic scintillator with an MPPC}
\label{sec:plasticMPPC}
\begin{figure*}[h]
\begin{subfigure}[t]{0.5\linewidth}
\includegraphics[width=\linewidth]{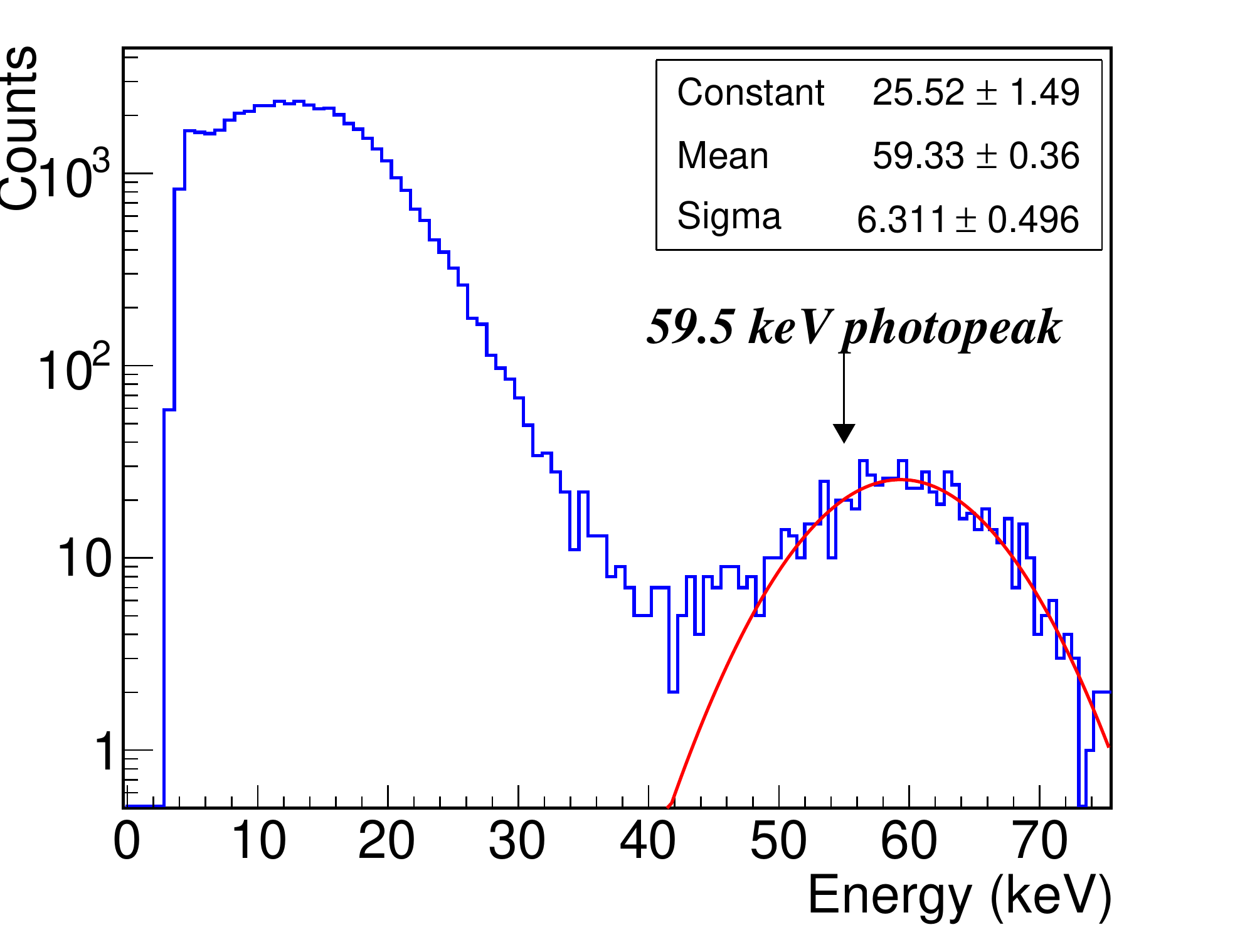}
\caption{\label{fig:PlasticHG}Am-241 spectrum using the HG amplifier of Citiroc.}
\end{subfigure}
\begin{subfigure}[t]{0.5\linewidth}
\includegraphics[width=\linewidth]{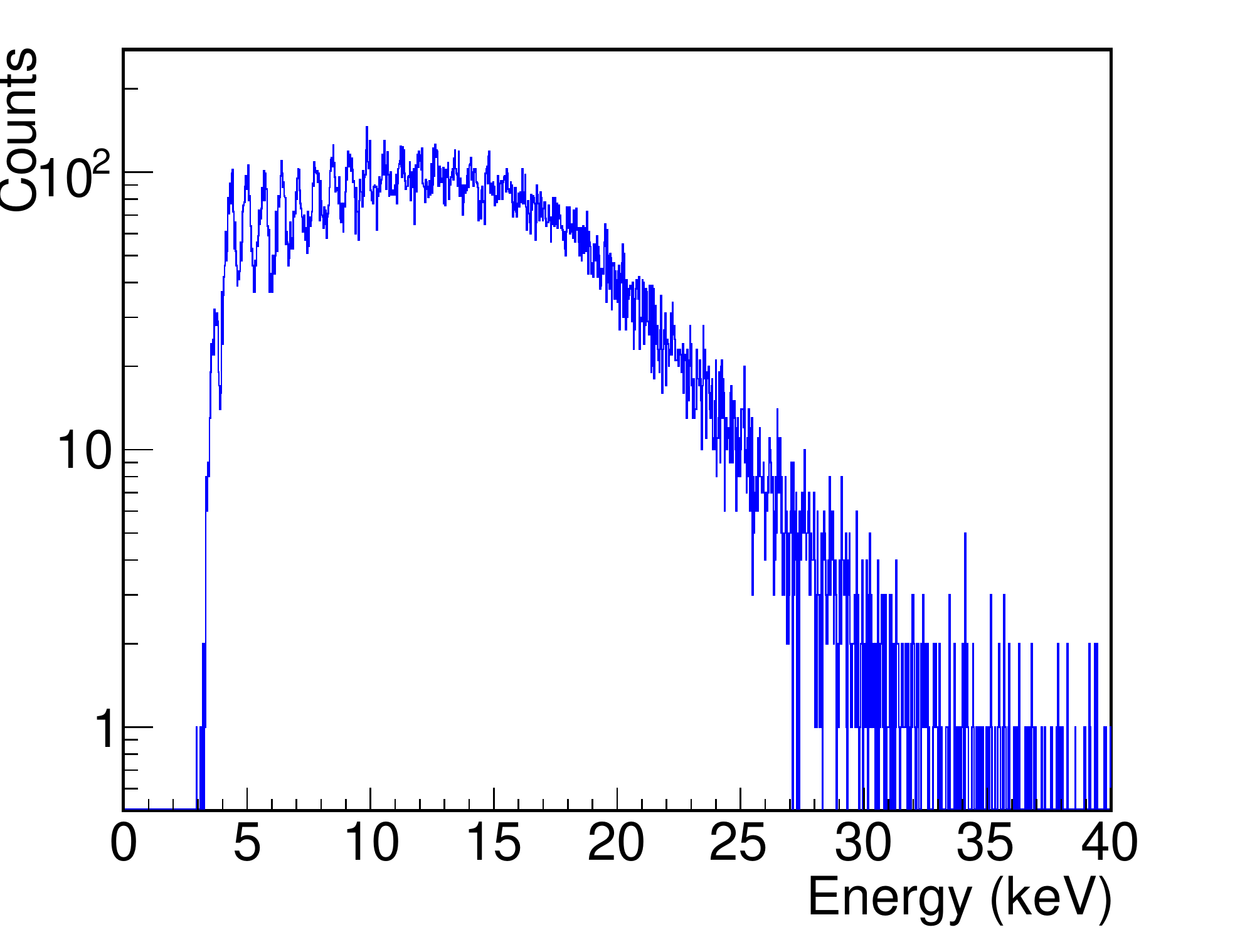}
\caption{\label{fig:Expanded} Expansion of the spectrum to show individual p.e. peaks.}
\end{subfigure}%
\caption{Energy spectrum of photons from Am-241 in the plastic scintillator coupled to an MPPC.}
\end{figure*}

The plastic scintillator is irradiated with photons from Am-241 (mainly 59.5 keV). A linear conversion (valid in the range indicated by the horizontal dashed lines in Figure~\ref{fig:LinearityHG}) is used to transform ADC counts to energy. The energy spectrum of Am-241 with a plastic scintillator is shown in Figure~\ref{fig:PlasticHG}. It shows a photoabsorption peak, fitted with a Gaussian function and a Compton continuum which is further expanded in Figure~\ref{fig:Expanded} to show the individual p.e. peaks obtained for the plastic scintillator. 

The ADC channel value corresponding to the Am-241 photoabsorption peak position is divided by the separation of consecutive p.e. peaks (seen in the spectrum) to estimate the light-yield. Under the given operating conditions, the light-yield of the plastic scintillator is \mbox{1.6 p.e./keV}. We note that the light-yield achieved in this configuration is higher than previous measurements \cite{compass} possibly due to wrapping of the scintillator with a high reflectance material, ESR. It could also be due to near perfect match of coupling area between the MPPC (photosensitive area) and the plastic scintillator (footprint area).

\begin{figure*}[h]
\begin{subfigure}[t]{0.5\linewidth}
\includegraphics[width=\linewidth]{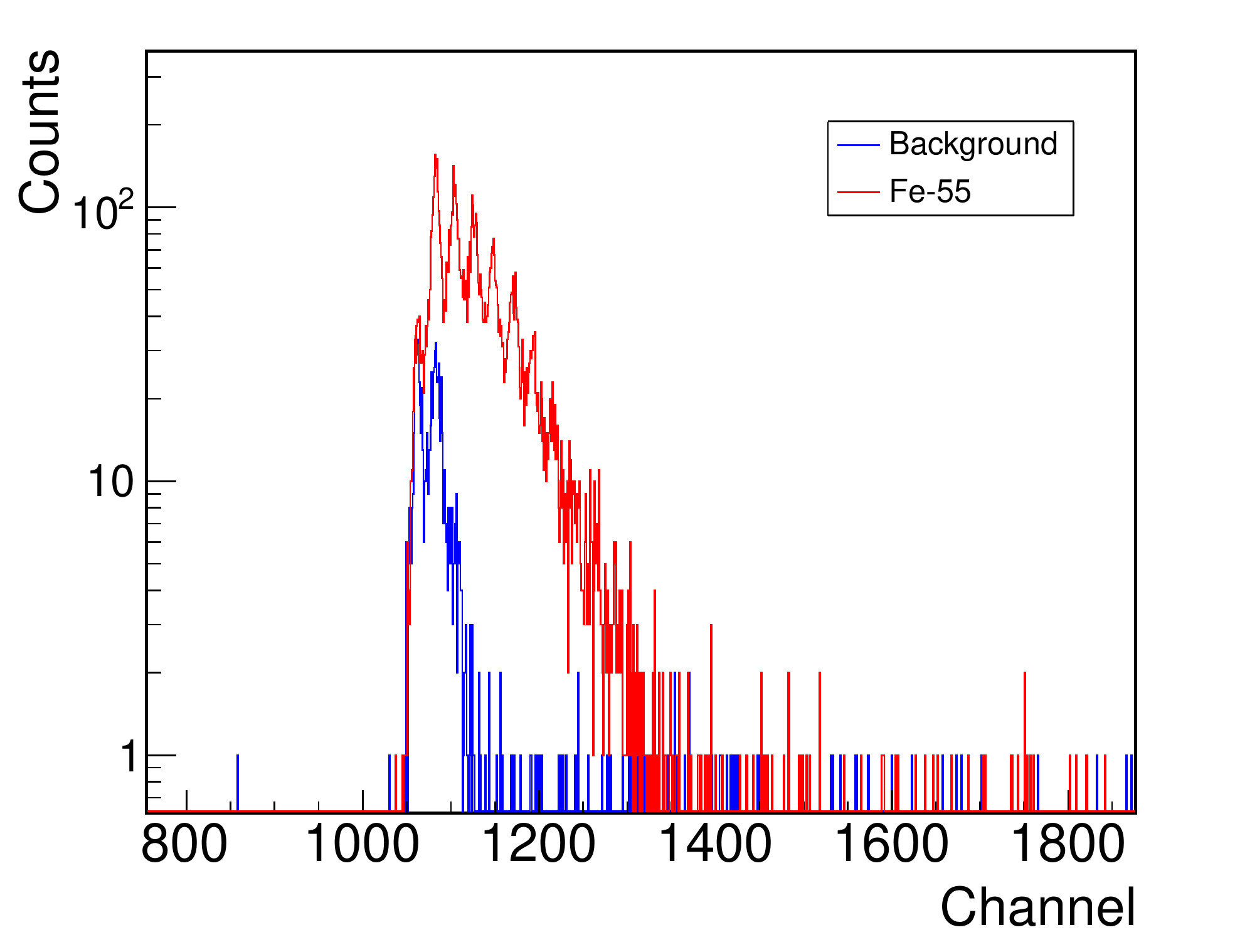}
\caption{\label{fig:iron} Fe-55 spectrum plotted with background. The individual p.e. peaks are also seen.}
\end{subfigure}
\begin{subfigure}[t]{0.5\linewidth}
\includegraphics[width=\linewidth]{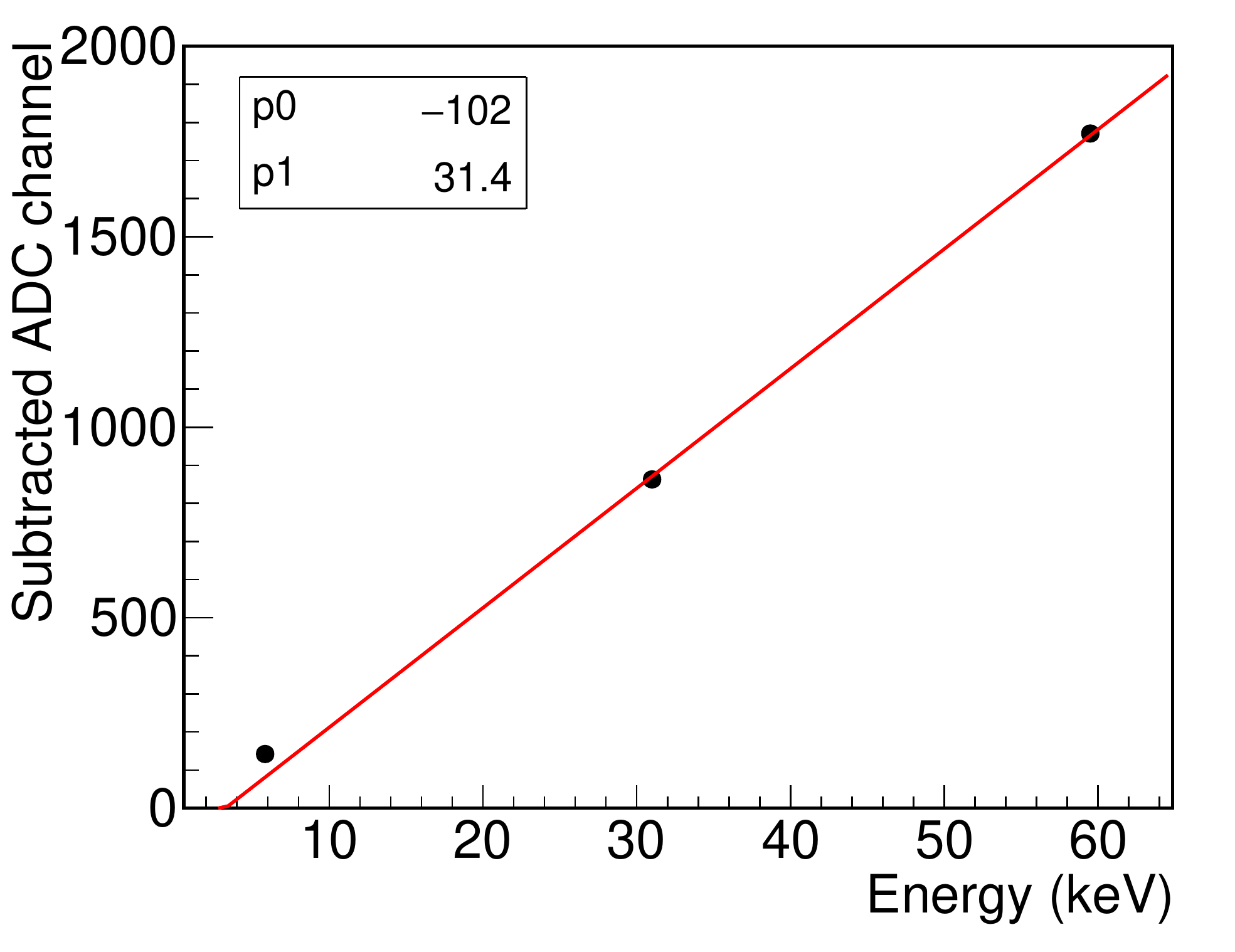}
\caption{\label{fig:enlin}Energy linearity of the plastic scintillator. The ADC channel on the vertical axis resulted from the subtraction of ASIC pedestal from the peak ADC channel. Error bars are smaller as compared to the data points.}
\end{subfigure}%
\caption{Response of the plastic scintillator coupled to an MPPC.}
\end{figure*}

The energy range, when the plastic scintillator is coupled to an MPPC, is estimated in the following manner. The lowest ADC channel (above threshold) and highest ADC channel in the linear range of the HG amplifier are calibrated with the known ADC channel of the Am-241 photoapbsorption peak. Assuming a linear response of the plastic scintillator, the estimated energy range is \mbox{4.5 keV$-$68 keV}. The non-linear response of EJ-248M was studied by the POLAR collaboration and Birk's constant was determined \cite{nonlinearity}. The value of Birk's constant cannot be used directly to correct the light yield in our work due to the different size of the plastic scintillator. Instead, in order to confirm the low energy detection limit in the plastic scintillator, we irradiate it with photons from Fe-55 (5.9 keV). The Fe-55 spectrum is shown along with the background spectrum in Figure~\ref{fig:iron}. The photoabsorption peak is fitted with a Gaussian function to evaluate the lowest energy detected. We also studied the energy linearity with the 6 cm long plastic scintillator and the result is shown in Figure~\ref{fig:enlin}. The three data points correspond to photons from Fe-55, Ba-133 and Am-241, from low to high energy, respectively. Data points are fitted with a linear function $\mathrm{(p0+p1\textit{x})}$ and the non-linearity is negligible in the energy range considered. 

\begin{figure}[h]
\centering 
\includegraphics[width=8cm, height=6cm]{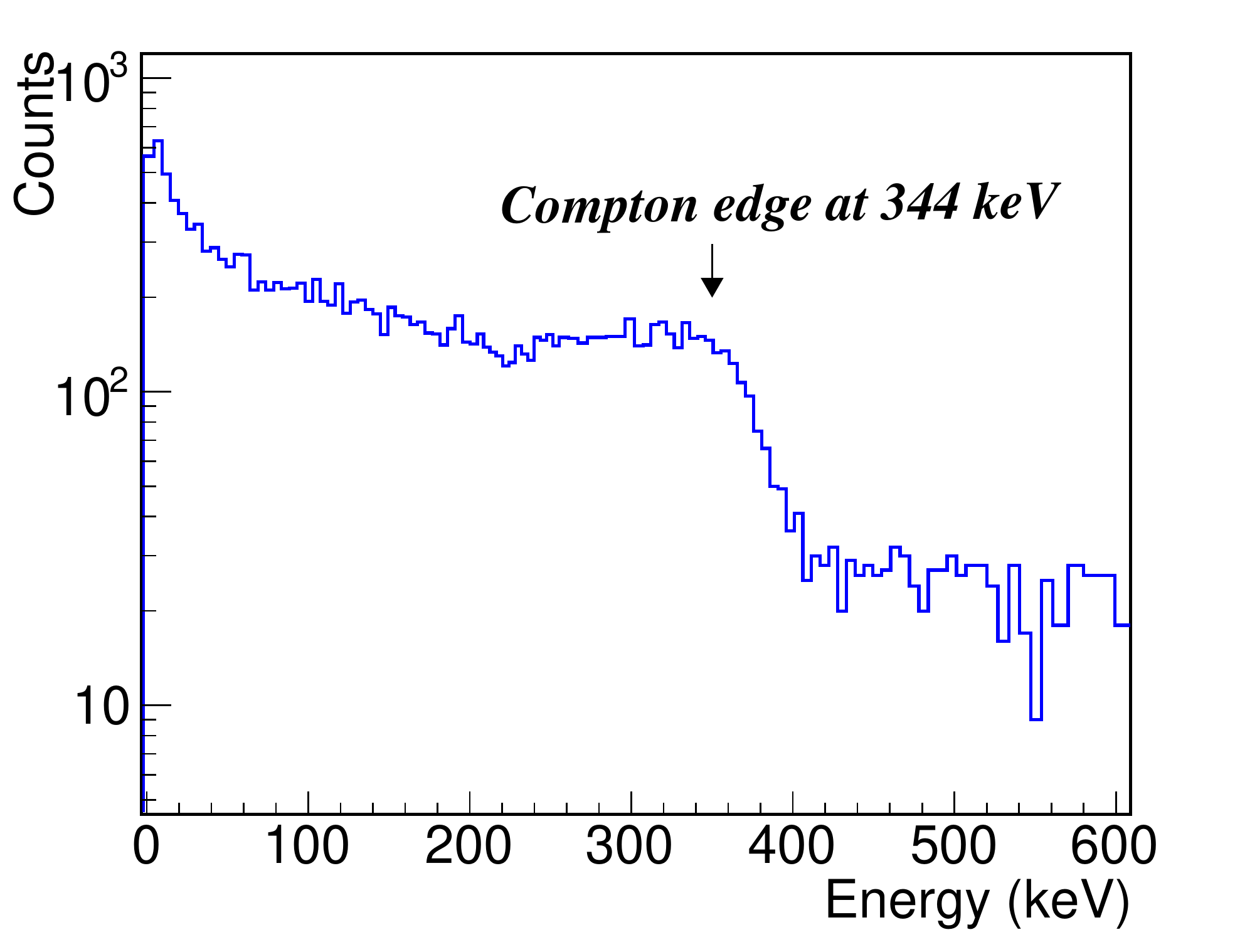}
\caption{\label{fig:PlasticLG}Na-22 spectrum for plastic scintillator using LG amplifier of Citiroc.}
\end{figure}

The plastic scintillator is also irradiated with photons from Na-22 (using 511 keV annihilation radiation) to estimate the energy range using the LG amplifier of the Citiroc. The spectrum is shown in Figure~\ref{fig:PlasticLG} which corresponds to the Compton-scattered photons. The energy range using the LG amplifier is 50 keV$-$400 keV (see Table~\ref{tab:energyrange}).

\subsubsection{GAGG scintillator with an MPPC}
\label{sec:GAGGMPPC}
\begin{figure*}[h]
\begin{subfigure}[t]{0.5\linewidth}
\includegraphics[width=\linewidth]{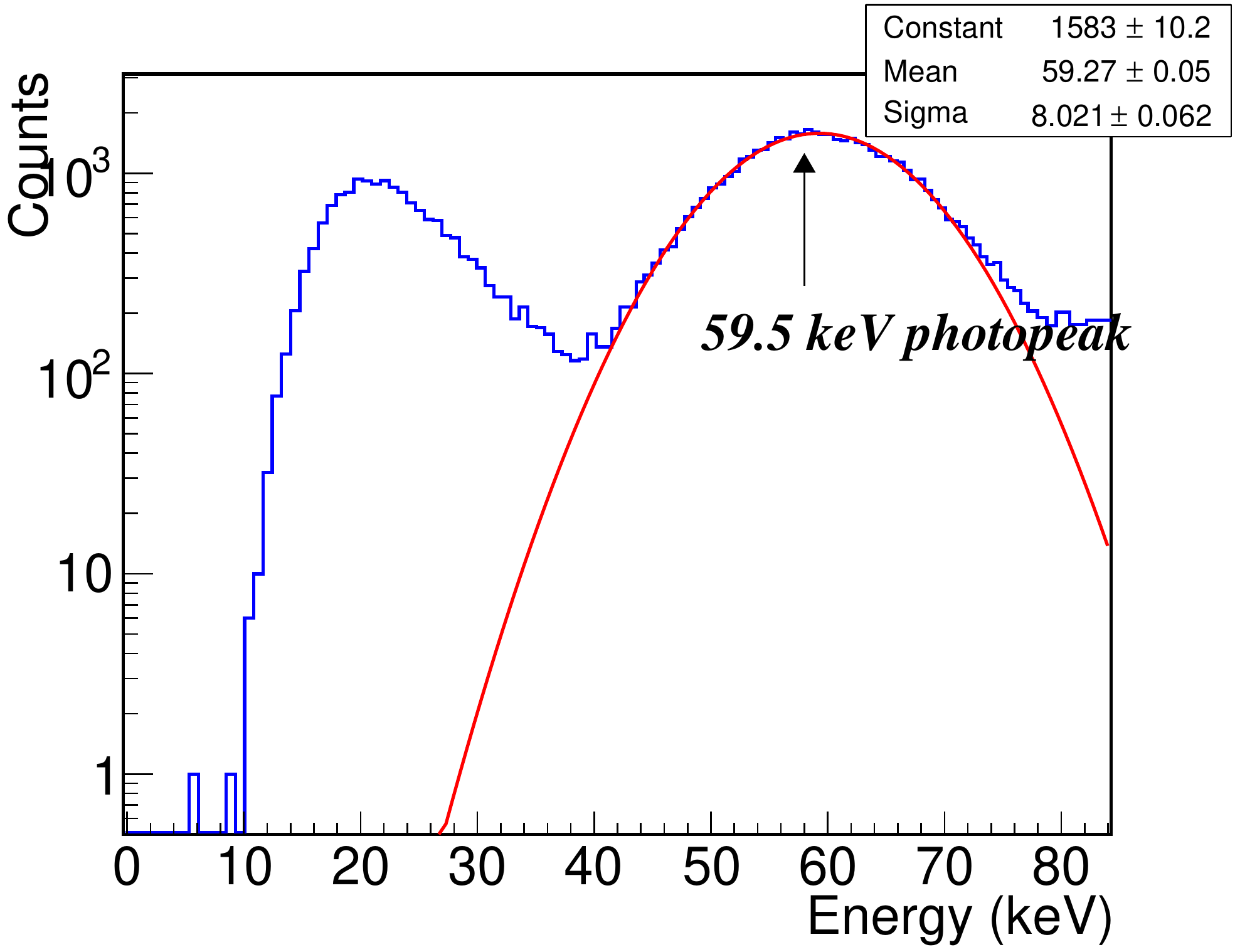}
\caption{\label{fig:HGenergy} Am-241 spectrum using the HG amplifier of Citiroc.}
\end{subfigure}
\begin{subfigure}[t]{0.5\linewidth} 
\includegraphics[width=\linewidth]{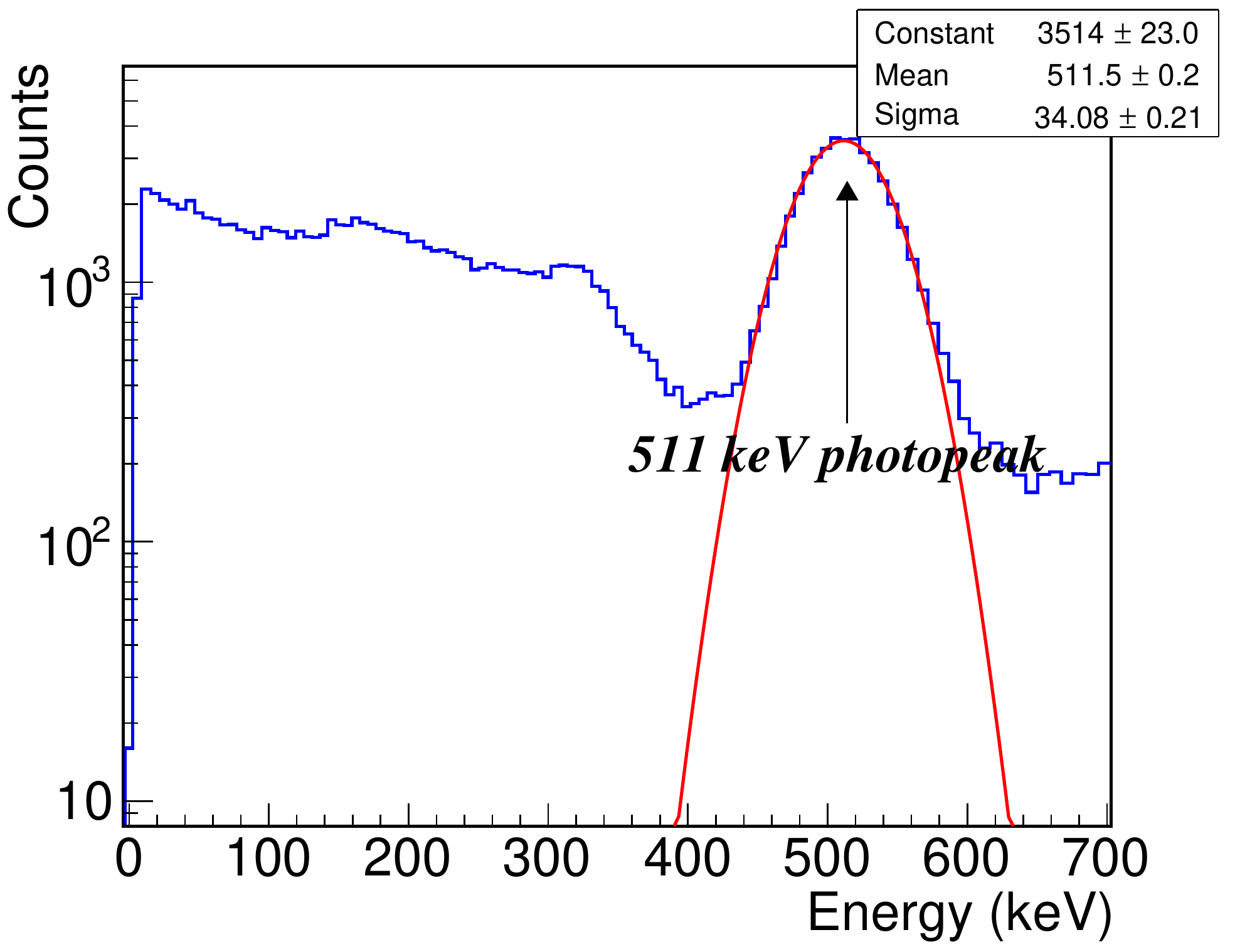}
\caption{\label{fig:LGenergy} Na-22 spectrum using the LG amplifier of Citiroc.}
\end{subfigure}%
\caption{Energy spectra of a GAGG scintillator coupled to an MPPC.}
\end{figure*}

The HG amplifier spectrum shown in Figure \ref {fig:HGenergy} is obtained by irradiating a GAGG scintillator with photons from Am-241. The right peak corresponds to photoabsorption of 59.5 keV photons while the left peak comprises low energy photons emitted by Am-241 and Compton events arising from 59.5 keV photons. The photoabsorption peak is fitted with a Gaussian function. In order to understand the behaviour of the GAGG scintillator with higher energy photons, it is irradiated using Na-22. Figure \ref{fig:LGenergy} shows the spectrum acquired using the LG amplifier of Citiroc. It consists of a Compton continuum and a photoabsorption peak at 511 keV (fitted with a Gaussian function).

All four GAGGs have been characterised and their performance is tabulated in Table~\ref{tab:performance}. As seen from the table, GAGG-D shows a lower light-yield as compared to other GAGGs. It is also seen that the estimated light-yield of the GAGG scintillators is lower than that of the plastic scintillator in the current set-up owing to the read-out geometry. The photosensitive area of the MPPC matches with that of the plastic scintillator while in case of the GAGG scintillator, only a small fraction of one edge is covered by the MPPC (see Figures~\ref{fig:schematic} and~\ref{fig:plasgagg}).  
\begin{table}[h]
\centering
\caption{Performance of the GAGG scintillators.}
\label{tab:performance}
\begin{tabular}{|l|c|c|}
\hline
					& {\bf Energy resolution (\%) } & {\bf Light-yield }  \\
					& {\bf at 59.5 keV} & {\bf (p.e./keV)}\\
\hline
GAGG-A & 35.9 $\pm$ 0.4 & 1.51 \\
GAGG-B & 37.9 $\pm$ 0.5 & 1.29 \\
GAGG-C & 38.7 $\pm$ 0.3 & 1.21 \\
GAGG-D & 42.5 $\pm$ 0.5 & 0.84\\
\hline
\end{tabular}
\end{table}

The spectra obtained from Citiroc, using HG and LG amplifiers, with both plastic and GAGG scintillators, are used to evaluate the energy range for X-ray detection. It is calculated in the linear range of Citiroc and is tabulated in Table \ref {tab:energyrange} for both types of scintillators. The functionality of the Citiroc suits the requirements of a hard X-ray (50$-$600 keV) polarimetric mission such as SPHiNX. The low energy threshold obtained when plastic is read-out with an MPPC is sufficient to allow detection of scattered 50 keV photons. This is explored further in next section.  

\begin{table}[h]
\centering
\caption{\label{tab:energyrange} Achieved energy range for plastic and GAGG scintillators with Citiroc.}
\smallskip
\begin{tabular}{|c|c|c|}
\hline
Amplifier&Plastic&GAGG\\
\hline
HG &5-68 keV &10-70 keV\\
\hline
LG & 50-400 keV  & 55-640 keV\\
\hline
\end{tabular}
\end{table}

\section{X-ray polarimetry with a plastic scintillator coupled to an MPPC}
 
\label{sec:xraypolarimetry}
\subsection{Experimental set-up}
\begin{figure*}[h]
\begin{subfigure}[t]{0.4\linewidth}
\includegraphics[width=\linewidth]{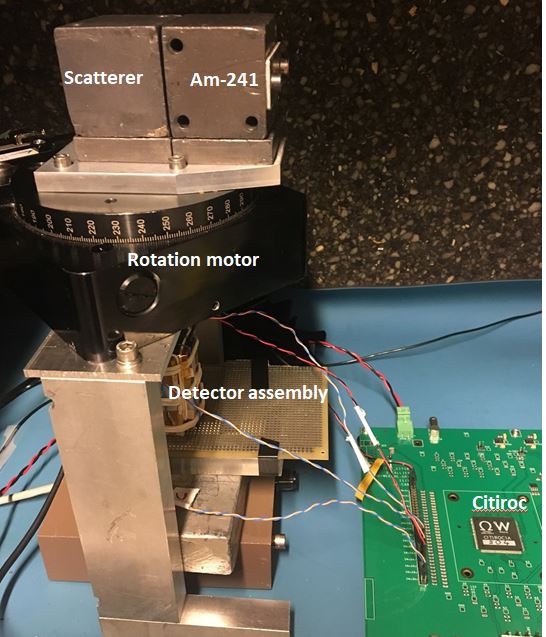}
\caption{\label{fig:Polarised} A photo of the test set up.}
\end{subfigure}
\begin{subfigure}[t]{0.5\linewidth}
\includegraphics[width=\linewidth]{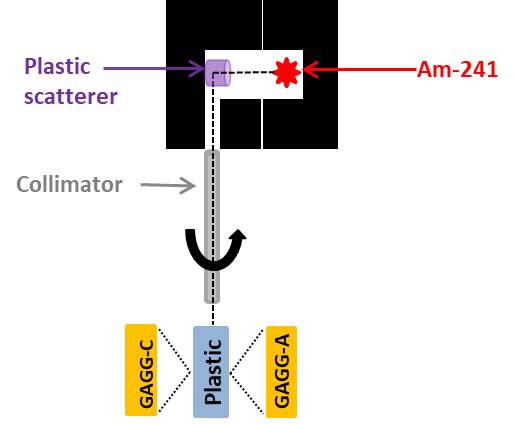}
\caption{\label{fig:crosssection} Schematic view of the test set-up.}
\end{subfigure}%
\caption{Experimental set-up for the coincidence measurement.}
\end{figure*}

\begin{figure}[h]
\centering 
\includegraphics[width=6cm, height=5cm]{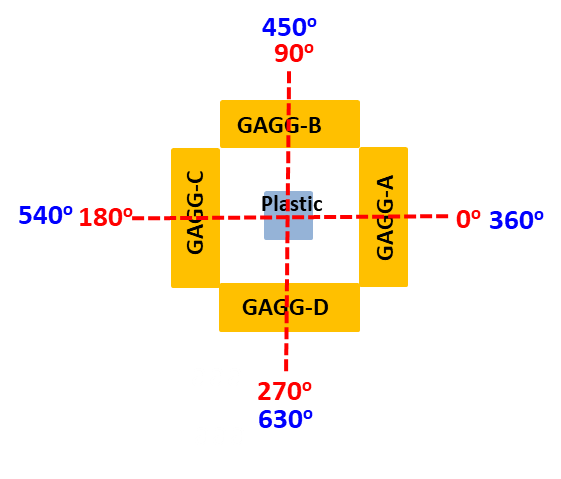}
\caption{\label{fig:placing}Placement of the scintillators and angular definitions.}
\end{figure}

X-ray polarimetry using Compton scattering requires recording coincident events in the plastic scintillator and one of the GAGG scintillators. The experimental set-up shown in Figure ~\ref{fig:Polarised} is used to make the coincidence measurement. It consists of an Am-241 source, scatterer, rotation motor, collimator, detector array and the Citiroc ASIC mounted on the evaluation board. The polarised radiation is obtained by scattering X-rays from Am-241 via a plastic scatterer. The radiation scattered at $90^{\circ}$ is nearly 100$\%$ polarised for $\sim$53 keV photons~\cite[Section 6.1 of Ref.][]{mozsi} and is collimated to irradiate the plastic scintillator as depicted in Figure~\ref{fig:crosssection}. The collimation results in a spot-size of $\sim$5 mm comparable to the plastic scintillator cross-section of 6 $\times$ 6 mm$^{2}$. By removing the scattering block, the plastic scintillator can be irradiated with unpolarised X-rays from the Am-241 source. The source assembly is mounted on the rotation table so as to rotate the source with respect to the scintillators. Figure~\ref{fig:placing} shows the placement of scintillators and the eight rotation angles \mbox{(0$^{\circ}-$ 630$^{\circ}$ at 90$^{\circ}$ intervals)} used in this study. For a polarized beam, a higher fraction of the radiation will be scattered into one pair of opposing GAGG scintillators compared to the other pair, depending on the angle of rotation of the source.

\subsection{Analysis and results}
The plastic scintillator in the coincidence set-up is first irradiated with unpolarised (direct) photons from Am-241 for 3 minutes. The spectrum is acquired at each of the eight rotation steps. This test is repeated with the polarised photons but the time of acquisition is 20 minutes for each rotation angle because of the lower rate of scattered photons. Coincident events are characterised by a GAGG interaction labeled in the Citiroc output file with "hit flag=1" and an associated plastic energy deposit above a predefined threshold (ADC value $>$ 1050 corresponding to $\sim$2.5 keV).   

\begin{figure*}[h]
\centering 
\includegraphics[width=\textwidth]{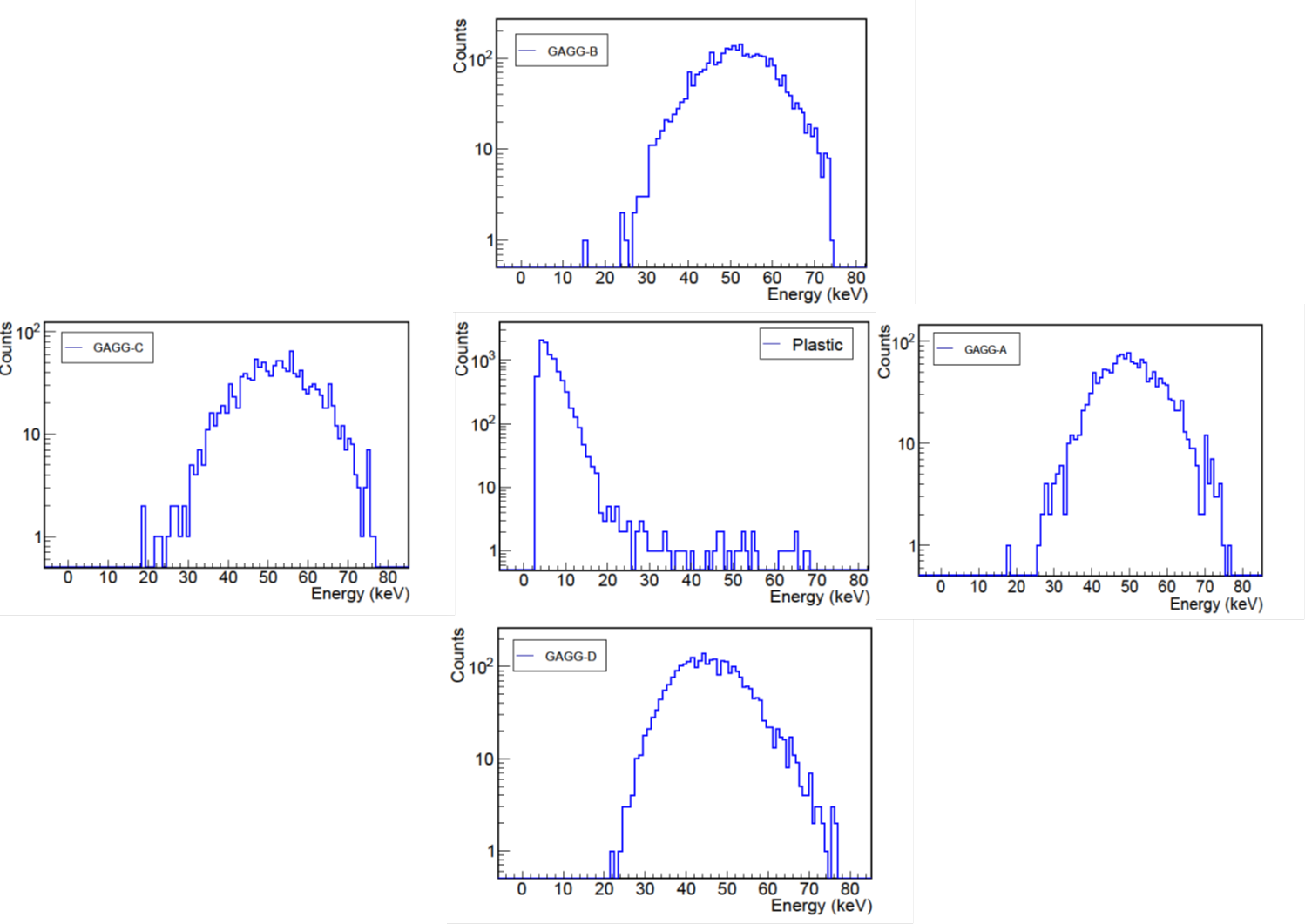}
\caption{\label{fig:polspectra} Coincidence spectra for  each of the five scintillators. The positions of the plots reflect the relative placement of the corresponding scintillators.}
\end{figure*}

\begin{figure}[h]
\centering 
\includegraphics[width=8cm, height=6cm]{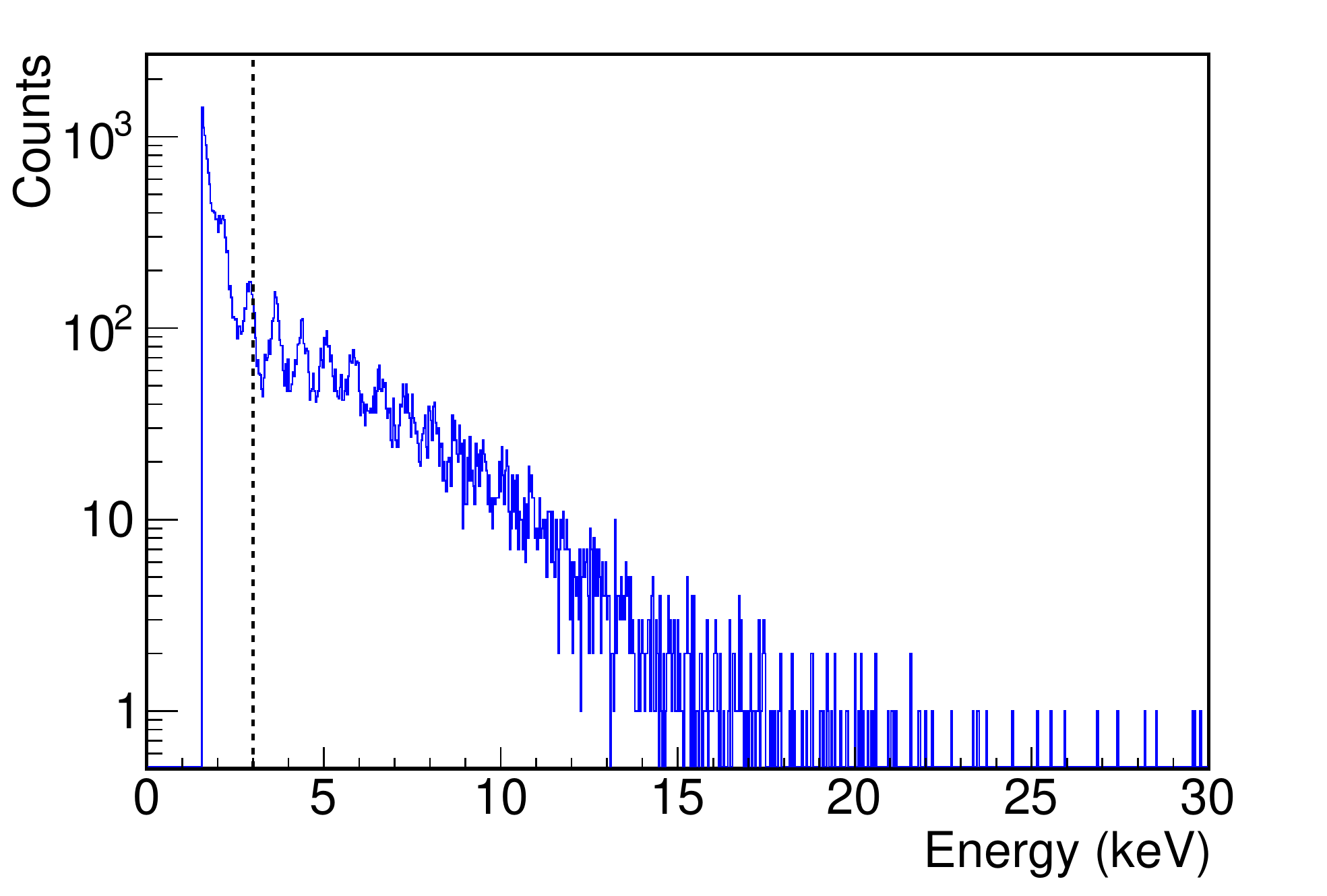}
\caption{\label{fig:polarisedplastic}Low energy spectrum with plastic scintillator in the coincidence measurement. The vertical dashed line at 3 keV is added to guide the eye.}
\end{figure}

The coincidence spectra obtained after applying these selections on the events from the polarised source are shown in Figure~\ref{fig:polspectra}. The plastic scintillator spectrum shows a Compton continuum while the four GAGG scintillators each show a photoabsorption peak corresponding to $\sim$48 keV photons (scattering at on average 90$^{\circ}$ of $\sim$53 keV photons). The low-energy detection limit of the plastic scintillator using the coincidence technique reduces to $\lesssim$3 keV, as shown in Figure~\ref{fig:polarisedplastic}. This figure is obtained using a linear conversion of ADC channel to energy. Previous work \cite{gap2019} has also reported reduction in the energy threshold achievable for a plastic scintillator by using the coincident detection. Our measurements show that it is possible to achieve a low threshold even at room temperatures, compared to the thresholds at much lower temperatures as reported in previous measurements \cite{gap2019}.

\begin{figure*}[h]
\begin{subfigure}[t]{0.5\linewidth}
\includegraphics[width=\linewidth]{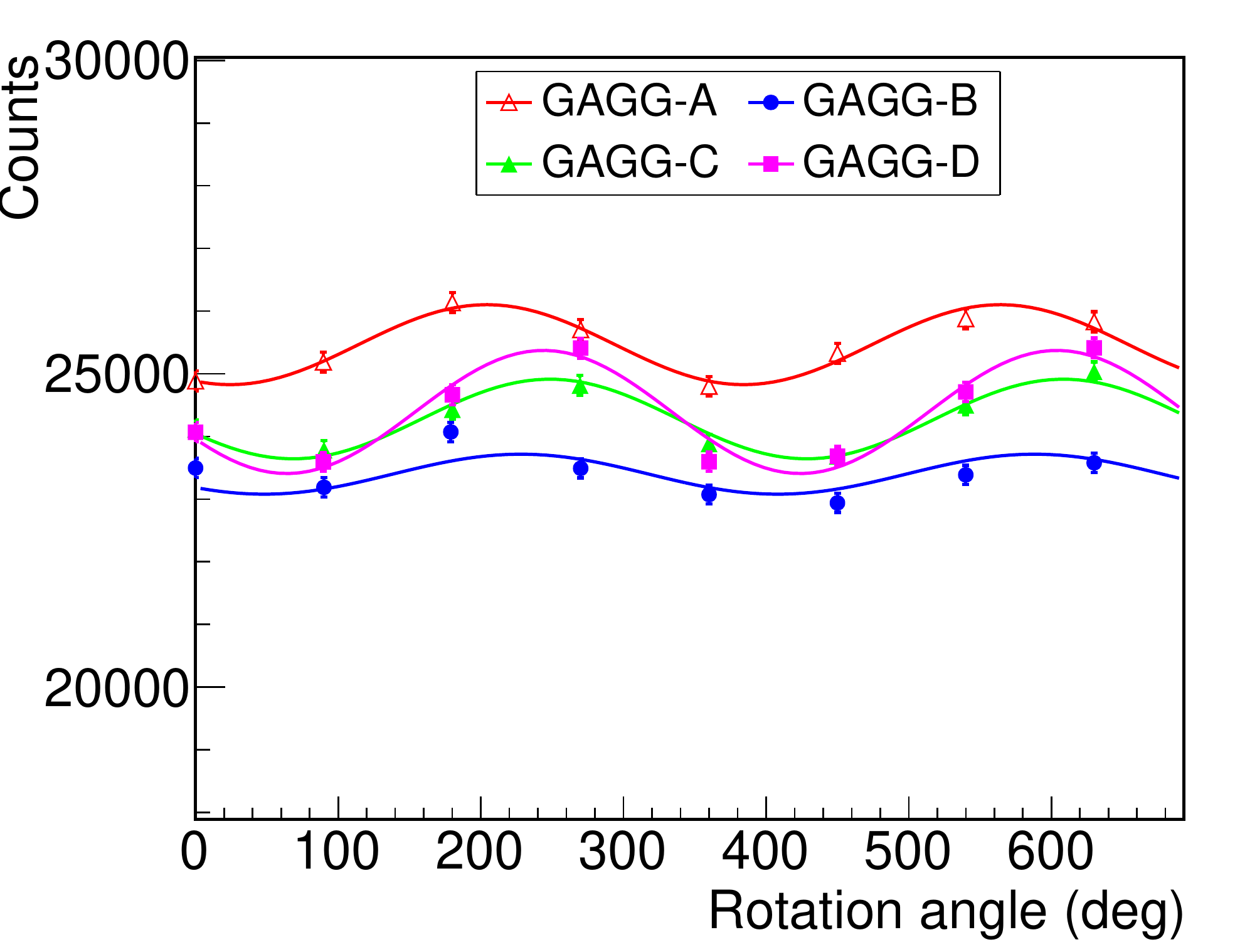}
\caption{\label{fig:unpolarised} Modulation curves obtained with an unpolarised beam. Error bars are small and partially obscured by the data points.}
\end{subfigure}
\begin{subfigure}[t]{0.5\linewidth}
\includegraphics[width=\linewidth]{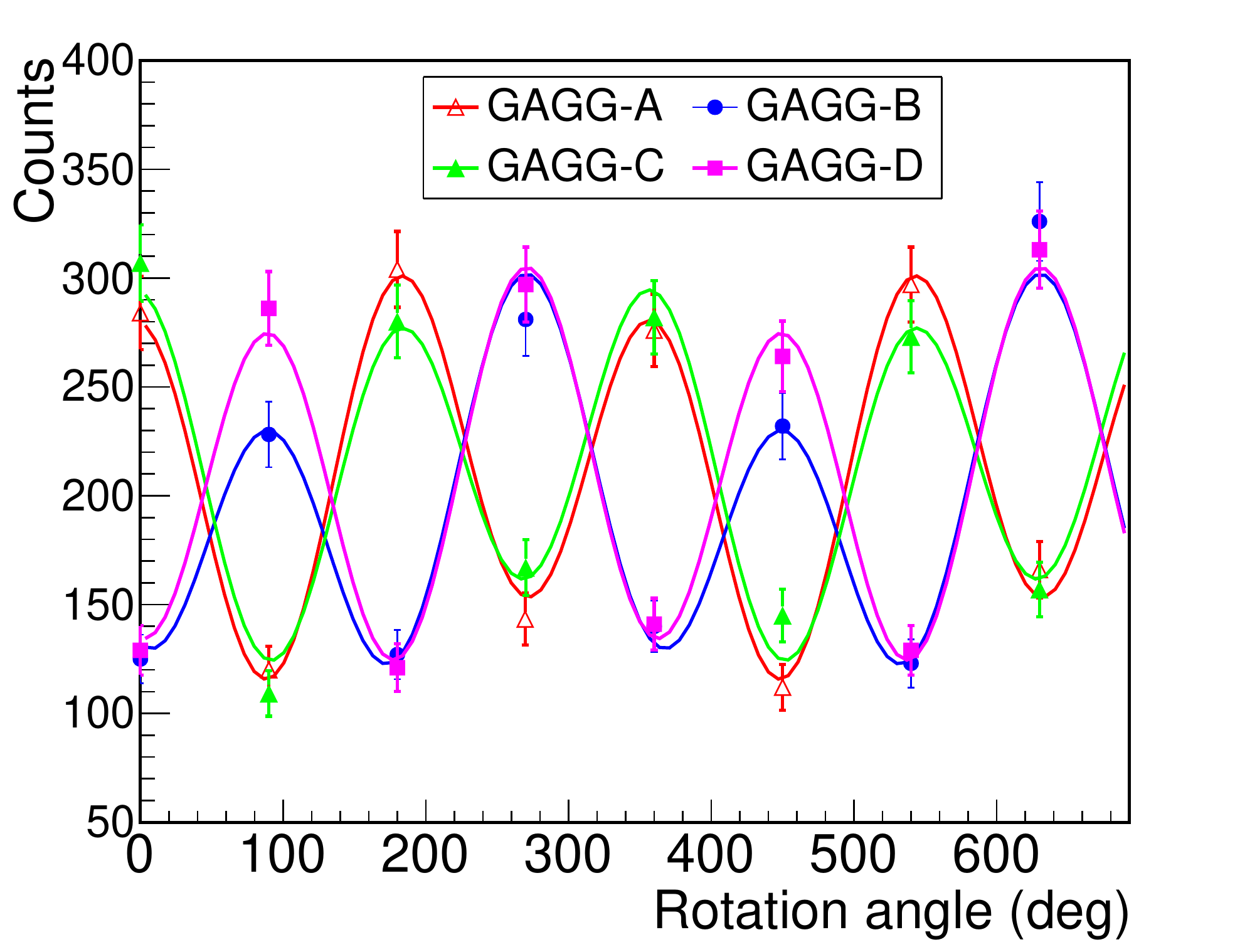}
\caption{\label{fig:Modcurve} Modulation curves obtained with a polarised beam.}
\end{subfigure}%
\caption{Modulation curves of four GAGG scintillators.}
\end{figure*}

The spectrum of each GAGG scintillator (shown in Figure~\ref{fig:polspectra}) is integrated over an ADC range (1100$-$3600) corresponding approximately to 20$-$70 keV in order to obtain the total counts. Integrated counts for the four GAGG scintillators are plotted as a function of rotation angle. \mbox{Figure~\ref{fig:unpolarised}} shows the modulation curves obtained with the unpolarised incident beam. All GAGG scintillators are expected to have uniform counts at all rotation angles. However, there exists a small residual variation which could arise from any of the reasons mentioned below.
\begin{itemize}
\item Plastic scintillator not fully centered within the GAGG scintillator array (scintillator offset)
\item Beam not fully centered on the plastic scintillator (beam offset) 
\item Photon beam incident at an angle with respect to the plastic scintillator (beam off-axis)
\end{itemize}

Each of these curves is fitted with a constant and a sinusoid of 360$^{\circ}$ periodicity, $\mathrm{p0+p1\cos(\textit{x}+p2)}$ where p0 is a constant term, and p1, p2 are the amplitude and phase of the modulation. Conducting two revolutions with the source rotation gives eight data points per curve, allowing the use of additional components to account for the residual variation. The addition of the sinusoidal function improves the reduced $\chi^2$ over a constant fit alone. From the amplitude of the 360$^{\circ}$ component, it can be inferred that there is either an off-axis or off-center alignment, and from the phase of this component, it can be inferred that in which direction the offset is. From the fitting parameters, the offset/misalignment can be quantified through simulations, as exemplified in [24]. The 360$^{\circ}$ component amplitude (p1) is only 3\% of the constant term (p0), indicating that these effects are very small. 

For a polarised beam, the modulation in counts with respective errors are plotted in Figure~\ref {fig:Modcurve}. The modulation curves are fitted with a function, $\mathrm{p0+p1\cos(\textit{x}+p2)+p3\cos(2\textit{x})}$, where $\mathrm{p3\cos(2\textit{x})}$ corresponds to the modulation due to polarisation. It is clearly seen that the polarised source leads to an increase of counts in opposing GAGGs and that the maximum of this asymmetry moves with the source rotation, as expected.

For reference, the modulation factor is estimated by using the fit parameters obtained for individual modulation curves. The average modulation factor obtained from the four GAGG scintillators is \mbox{0.36}. This is an order of magnitude greater than the 3\% variation arising from the offset/misalignment. Furthermore, the polarization signature is present in the 180$^{\circ}$ component of the asymmetry, which is independent from the 360$^{\circ}$ component mentioned above. The distinct modulation of counts demonstrates feasibility of the coincidence measurement with the plastic scintillator read out with an MPPC.

\section{Conclusion and discussion}

Our results demonstrate that MPPCs used in a Citiroc-based read-out system have the potential to replace PMTs in a hard X-ray polarimeter. The set-up used here comprises a 60~mm long, 6 $\times$ 6~mm$^2$ plastic scintillator bar is surrounded by 27.5~mm wide slabs of GAGG(Ce) scintillator. The MPPC sensitive area, 6 $\times$ 6~mm$^2$, is well matched to that of the plastic-scintillator bar resulting in a light-yield of 1.6 p.e./keV and, consequently, a detection threshold of $\sim$5~keV at room temperature. This allows the detection of $\sim$50~keV X-rays scattering through a polar angle of 90$^\circ$ in the plastic scintillator. A lower threshold should be possible if the MPPC is operated at a lower temperature, for example using a Peltier element. Polarimetry events are identified as coincident hits in the plastic and GAGG scintillators. This mitigates the effect of MPPC dark-noise and reduces the low-energy detection threshold. For precisely determining the low-energy limit, the non-linear energy response of plastic scintillator needs to be thoroughly studied e.g. using a synchrotron beam. The use of GAGG(Ce) provides good energy resolution for scattered photons (37.9\% at 59.5 keV and 13.7\% at 511 keV), potentially allowing spectro-polarimetric observations of sources. 

In order to improve the modulation factor of our demonstrator design, the GAGG(Ce) scintillators could be segmented in length and width, optically isolated, and equipped with individual MPPC read-outs. In order to realise a large-effective-area polarimeter, e.g. as required for GRB polarimetry, a promising way forward is to use MPPC arrays where each MPPC element is equipped with a plastic or GAGG(Ce) scintillator. The layout of the scintillators should be optimised using computer simulations. For the SPHiNX design illustrated in Figure~\ref{fig:geo}, the large plastic scintillator scatterer could be read out by an MPPC array (e.g. Hamamatsu S13361-6050AE-04), with the individual cell output summed to form a single signal. Although foreseen primarily as a laboratory demonstrator, the current design is most suited to a focal plane polarimeter where X-ray optics provide the effective area required for sensitive observations. The X-Calibur mission \cite{xcalibur} uses an analogous approach where a passive beryllium scatterer is surrounded by pixelated CdZnTe semiconductor detectors.

The 32-channel Citiroc ASIC allows a compact read-out system to be realised. In the present demonstrator set-up, the Citiroc functionality is dictated by constraints posed by the Weeroc evaluation board. For example, this means that time-coincident events in the plastic and GAGG(Ce) scintillators are identified offline. This is not suitable for a space-based instrument where on-board storage and downlink capacity may be limited. In this case, coincidences would be identified in real-time using digital logic fed by the charge trigger outputs from the Citiroc. The logic could also examine signals from an anti-coincidence system before a decision is made regarding event registration. 

As well as fulfilling functional requirements, polarimeter components must operate reliably in the space environment. The low energy threshold can increase due to radiation induced enhancement of dark current  in MPPCs. Several groups are currently studying how to mitigate these effects. Detailed irradiation studies are required before MPPCs can be adopted for future space-based instruments. 
  
\section*{Acknowledgements}
This work is funded by the Swedish National Space Agency (grant numbers 232/16) and the Swedish Research Council (grant number 2016-04929). We thank Merlin Kole, University of Geneva, for providing the plastic scintillator used in this work. Hiromitsu Takahashi, Hiroshima University, is thanked for his assistance in procuring the GAGG scintillators. The work done by Masters' course students, Love Eriksson and Bill Burrau is gratefully acknowledged. Lastly, we would like to thank Linda Eliasson for helping with the rotation motor set-up.



\bibliographystyle{elsarticle-num} 
\bibliography{citations}


%
%
%
\end{document}